\documentclass[prl,twocolumn]{revtex4-1}

\usepackage{epsfig,amsmath}
\usepackage{subfigure}
\usepackage{graphicx}
\usepackage{dcolumn}
\usepackage{stmaryrd}
\usepackage{mathrsfs}
\usepackage{pifont}
\usepackage{amsthm}
\usepackage{amssymb}
\usepackage{bm}
\usepackage{latexsym}
\usepackage{color}
\usepackage{pdfpages}

\begin{document}

\title{Fast quantum algorithm for EC3 problem with trapped ions}
\author{Hefeng Wang$^1$}
\author{Lian-Ao Wu$^{2,3}$}
\affiliation{$^{1}$Department of Applied Physics, Xi'an Jiaotong University, Xi'an
710049, China\\
$^{2}$Department of Theoretical Physics and History of Science, The Basque
Country University~(EHU/UPV), P. O. Box 644, 48080 Bilbao, Spain \\
$^3$IKERBASQUE, Basque Foundation for Science, 48011 Bilbao, Spain}

\begin{abstract}
Adiabatic quantum computing~(AQC) is based on the adiabatic principle, where
a quantum system remains in an instantaneous eigenstate of the driving
Hamiltonian. The final state of the Hamiltonian encodes solution to the
problem of interest. While AQC has distinct advantages, recent researches
have shown that quantumness such as quantum coherence in adiabatic processes
may be lost entirely due to the system-bath interaction when the evolution
time is long, and consequently the expected quantum speedup dose not
show up. Here we propose a fast-signal assisted adiabatic quantum algorithm.
We find that by applying a sequence of fast random or regular signals during
the evolution process, the runtime can be reduced greatly, yet
advantages of the adiabatic algorithm remain intact. Significantly, we
present a \emph{randomized} Trotter formula and show that the driving
Hamiltonian and the sequence of fast signals can be implemented
simultaneously. We apply the algorithm for solving the $3$-bit exact cover
problem~(EC$3$) and put forward an approach for implementing the problem
with trapped ions. \noindent
\end{abstract}

\maketitle

\emph{Introduction.}-- The adiabatic principle addresses quantum evolution
governed by a slowly-varying Hamiltonian where the system will stay near its
instantaneous ground state \cite{Born,Messiah}. It has a variety of
applications in quantum information processing, such as adiabatic quantum
computing~\cite{Farhi}, fault-tolerance against quantum errors~\cite{Childs}%
, and universal holonomic quantum computation~\cite{Aharonov,Zanardi,Carollo}
based on the Berry's phase~\cite{Berry,Zee,Barry}. AQC is one of quantum
computing models that have potential in solving certain problems much faster
than their classical counterparts, in particular factoring large integers~%
\cite{shor}, searching unsorted database~\cite{grover} and simulating
quantum many-body problems~\cite{Fey}.

Adiabatic quantum computation processes in a way where the quantum system
time-evolves from the ground state of an initial Hamiltonian to that of the
final Hamiltonian, which encodes the solution to the problem of interest. In
AQC, a long evolution time guarantees that the final state reaches the
ground state of the final problem Hamiltonian. This requires long coherence
time in experimental implementation of the process. As such, the evolution time is
crucial for AQC to be valid. If the evolution time is too long, quantumness
may become vanishingly small and consequently the quantum speedup over
classical computation will vanish. Recently an interesting experiment~\cite%
{lidar1} has been performed to address the crucial question: whether or not
a large-scale quantum device has the potential to outperform its classical
counterpart. The experimental test was done for finding the ground state of
an Ising spin glass model on the $503$-qubit D-Wave Two system which are
designed to be a physical realization of quantum annealing using
superconducting flux qubits. There was no evidence found for quantum
speedup. One of main reasons could be that the runtime is so long that
before the end of an adiabatic quantum algorithm, decoherence has completely
ruined its quantumness.

Because of decoherence, a quantum algorithm with short runtime is always
desired to keep its quantumness in the computational process. This is
particularly important for adiabatic quantum algorithms since a strict
adiabatic process requires an infinite runtime. In this paper, we present an
approach that speeds up AQC substantially by applying fast signals in the
dynamical evolution process, while keeping the high fidelity between the
final state and the eigenstate of the problem Hamiltonian. This approach is
experimentally feasible for implementing adiabatic quantum algorithms. We
demonstrate this approach by solving a $3$-bit exact cover problem~(EC$3$).

\emph{The Algorithm.}-- The EC$3$ problem is a particular instance of
satisfiability problem and is one of the NP-complete problems. No efficient
classical algorithm has been found for solving this problem. On a quantum
computer the EC$3$ problem can be formulated as follows~\cite{farhi1, Farhi}%
: for a Boolean formula with $M$ clauses
\begin{equation}
C_{1}\wedge C_{2}\wedge \cdots \wedge C_{M}\text{,}
\end{equation}%
where each clause $C_{l}$ is true or false depending on the values of a
subset of the $n$ bits, and each clause contains three bits. The clause is
true if and only if one of the three bits is $1$ and the other two are $0$.
The task is to determine whether one~(or more) of the $2^{n}$ assignments
satisfies all of the clauses, and find the assignment(s) if it exists.

In Ref.~\cite{farhi1,Farhi}, a quantum adiabatic algorithm for solving the EC%
$3$ problem has been proposed. In this algorithm, the time-dependent
evolution Hamiltonian $H_{0}\left( t\right) $ is
\begin{equation}
H_{0}\left( t\right) =J_0\left[\left( 1-t/T\right)
H_{B}+\left(t/T\right)H_{P}\right],
\end{equation}%
where $H_{B}$ is the initial Hamiltonian whose ground state is used as the
initial state, $H_{P}$ is the Hamiltonian of the EC$3$ problem whose ground
state is the solution to the EC$3$ problem and $T$ is the total evolution
time or the runtime. Here $J_0$ is the strength of the Hamiltonian and is
set as $J_0 =1$ in this paper. In this algorithm, the Hamiltonian of the
system evolves adiabatically from $H_{B}$ to the problem Hamiltonian $H_{P}$%
, meaning that the system evolves from the ground state of $H_{B}$ to the
ground state of $H_{P}$. $H_{B}$ is defined as
\begin{equation}
H_{B}=\sum_{C}H_{B,C}.
\end{equation}%
where $H_{B,C}$ is the Hamiltonian of clause \textit{C}. Let $i_{C}$, $j_{C}$
and $k_{C}$ be the $3$ bits associated with clause \textit{C}. $H_{B,C}$\ is
defined as
\begin{equation}
H_{B,C}=H_{B}^{i_{C}}+H_{B}^{j_{C}}+H_{B}^{k_{C}},
\end{equation}%
with
\begin{equation}
H_{B}^{i}=\frac{1}{2}\left( 1-\sigma _{x}^{i}\right)
\end{equation}%
and $\sigma _{x}^{i}$\ are the Pauli matrices. The Hamiltonian $H_{P}$ for
the EC$3$ problems is defined as follows: for each clause \textit{C}, one
can define an \textquotedblleft energy\textquotedblright\ function
\begin{equation}
h_{C}(z_{i_{C}},z_{j_{C}},z_{k_{C}})\!=\!\Bigg\lbrace%
\begin{array}{c}
\!\!\!\!0,\,\,\mathrm{if}\,(z_{i_{C}},z_{j_{C}},z_{k_{C}})\ \mathrm{%
satisfies\ clause\ \emph{C}} \\
\!\!\!\hskip.0003in1,\,\,\mathrm{if}\,(z_{i_{C}},z_{j_{C}},z_{k_{C}})\
\mathrm{violates\ clause\ \emph{C}}\,%
\end{array}%
\end{equation}%
such that
\begin{equation}
H_{P,C}|z_{1}z_{2}\cdots z_{n}\rangle =h_{C}\left(
z_{i_{C}},z_{j_{C}},z_{k_{C}}\right) |z_{1}z_{2}\cdots z_{n}\rangle ,
\end{equation}%
where $|z_{j}\rangle $ is the $j$-th bit and has a value $0$ or $1$. Define
\begin{equation}
H_{P}=\sum_{C}H_{P,C},
\end{equation}%
and we then have $H_{P}|\psi \rangle =0$, if and only if $|\psi \rangle $ is
in a superposition of states $|z_{1}z_{2}\cdots z_{n}\rangle $, where the
bit string $z_{1}z_{2}\cdots z_{n}$ satisfies all of the clauses.

In what follows we will describe our approach for solving the EC$3$ problem
by applying a sequence of fast pulses during the dynamical process~\cite%
{Jun14}. We consider a Hamiltonian -- a \emph{dressed} $H_{0}(t)$,
\begin{equation}
H\left( t\right) =\left( 1+c\left( t\right) /J_{0}\right) H_{0}\left(
t\right) ,
\end{equation}%
where $c(t)$ represents a sequence of fast signals. Ref.~\cite{Jun14} shows
that the adiabaticity can be enhanced and even induced by $c(t)/J_{0}$ --
regular, random, and even noisy fast signals. Specifically, $c(t)/J_{0}$ could 
be a white noise signal in magnetic field, as exemplified in Ref.~\cite{Jun14}.  We will use this strategy to
speed up adiabatic quantum algorithms and then illustrate our general
approach by an experimentally feasible example.

We now come to explain the principle and experimental implementation of our
approach in terms of a simple but nontrivial EC$3$ problem. Consider a $4$%
-bit EC$3$ problem, where we select the $3$-bit set of clauses as \{$1,2,3$%
\}, \{$2,3,4$\}, and \{$1,2,4$\}. The solution to this problem is $%
|0100\rangle $. 

For this specific model, we show numerically that when $%
T_{0}>160$, the system enters the adiabatic regime and $F\approx0.999$ at $t=T$. In order to study the
contributions of fast signals, we set $T=40<T_{0}$ in the non-adiabatic
regime, and apply a sequence of fast regular pulses during the adiabatic
process. The interval between pulses is set as $0.08$, and the pulses
strengths are changed $s=0,0.5,1.0,2.0$, respectively. Fig.~$1$ shows the
dynamics of fidelity $F=|\langle \psi (t/T)|\psi
_{0}(t/T)\rangle |$ between the system wave function and the
instantaneous ground state of $H(t)$,  where $\psi (t/T)$ is the wave function governed by the
Schr\"{o}dinger equation or the time-ordering evolution operator and $|\psi
_{0}(t/T)\rangle $ represents the instantaneous ground state of the
Hamiltonian $H(t)$. It is clear in the figure that as the strength of pulses
increases, the adiabaticity is induced from a non-adiabatic regime and the
fidelity $F$ is approaching to one, in particular in the region where the solution is
encoded. The quality of pulse control can also be improved by increasing the
density of fast signals.

\begin{figure}[tbp]
\begin{minipage}{0.98\linewidth}
  \centerline{\includegraphics[width=0.98\columnwidth, height=0.4\columnwidth]{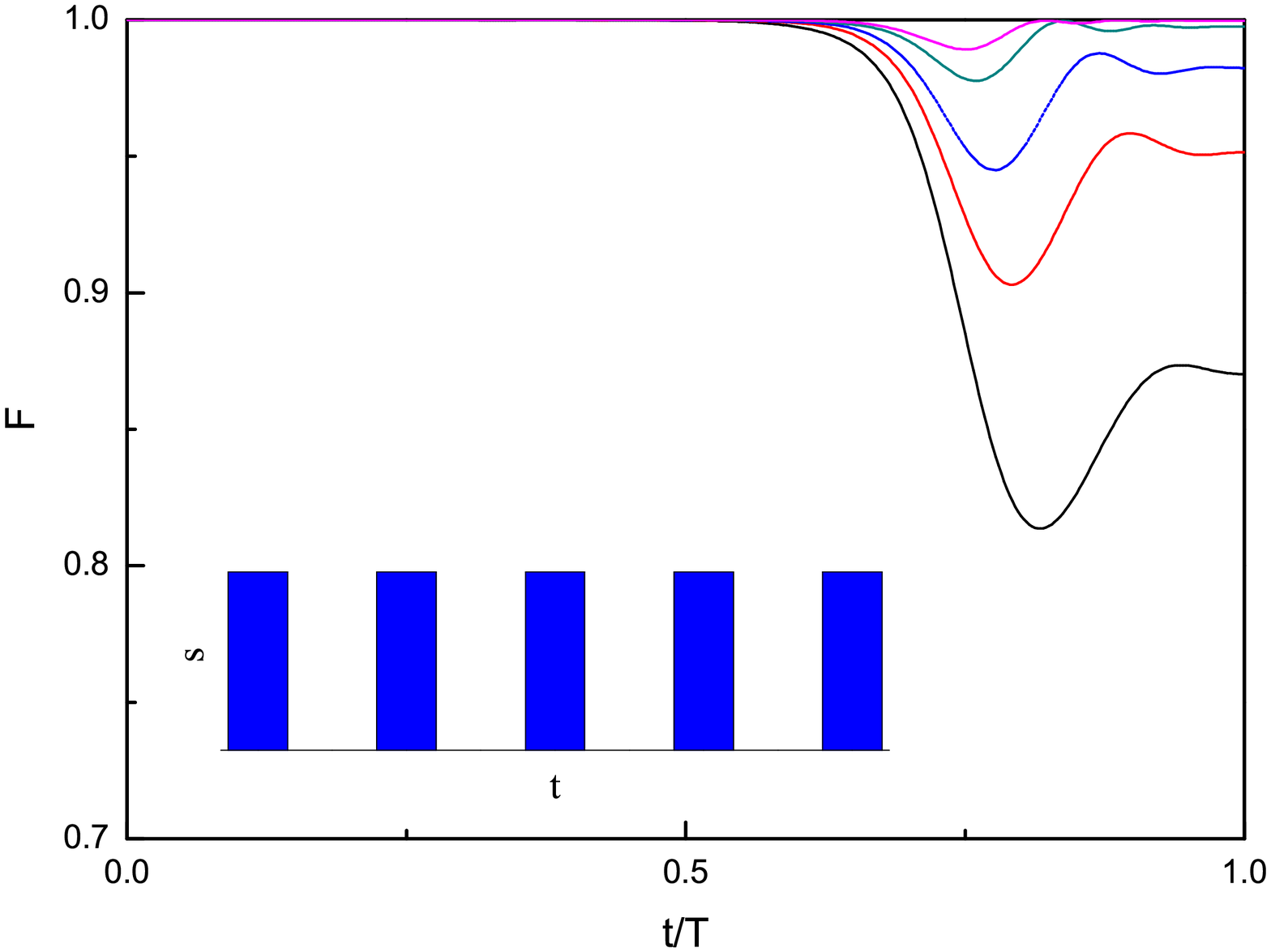}}
  \centerline{(1a)}
\end{minipage}
\vfill
\begin{minipage}{0.98\linewidth}
  \centerline{\includegraphics[width=0.98\columnwidth, height=0.4\columnwidth]{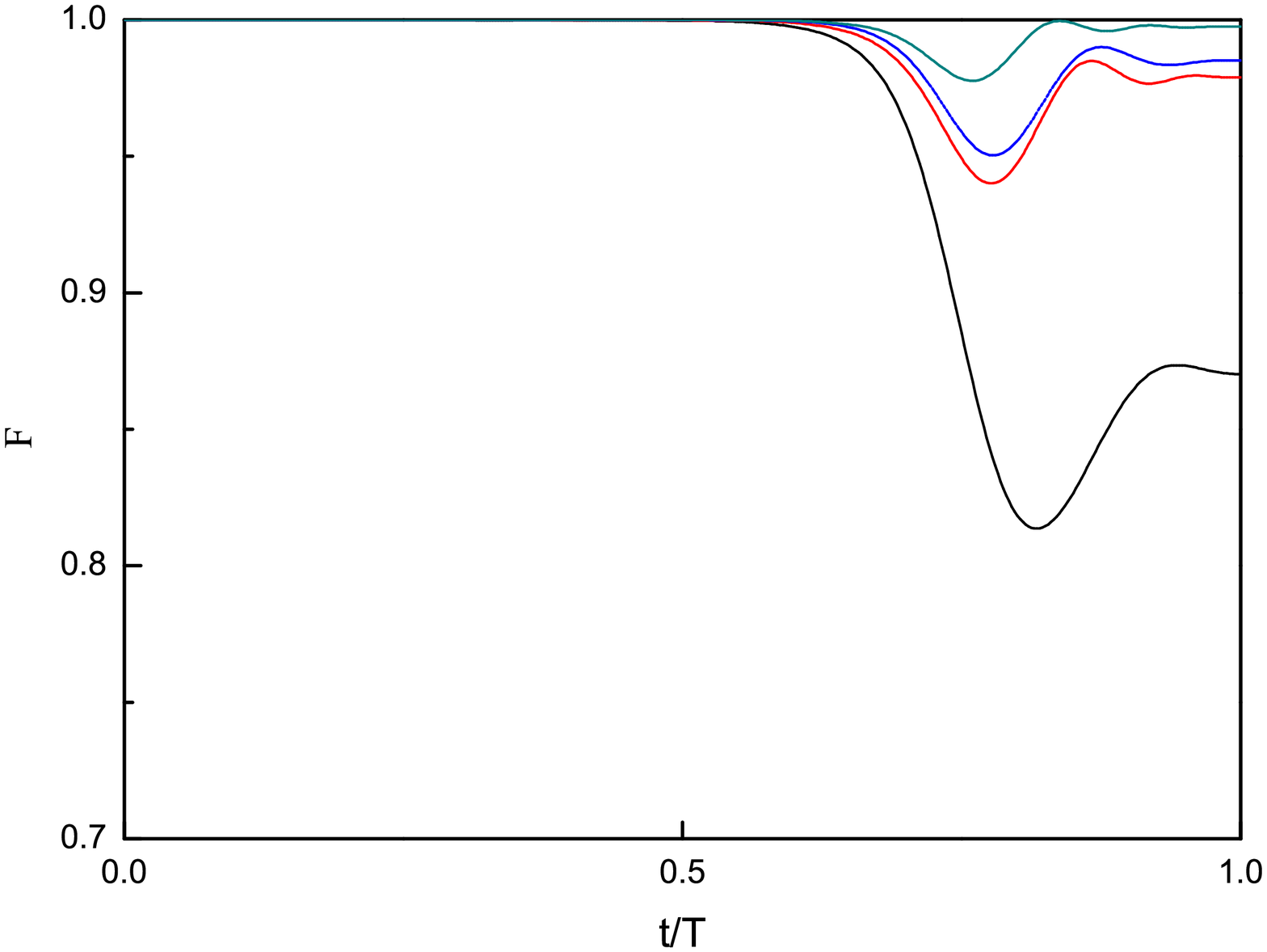}}
  \centerline{(1b)}
\end{minipage}
\caption{(Color online) Dynamics of fidelity $F=|\langle \protect\psi %
(t/T)|\protect\psi _{0}(t/T)\rangle |$. 
The evolution time or runtime $T=40$, the intervals between pulses are set
as $\Delta =0.08$. The black solid line shows the result for the strength $%
s=0$; the red dashed line for $s=0.5$; the blue dotted line for $s=1.0$ and the dash-dot dark cyan line shows for $s=2.0$. 
The magenta short dashed line shows the result of the evolution time $T_{0}=160$ when the
system {\em enters} the adiabatic regime justified by $F(T)\ge 0.999$. For this model we consider this curve as a reference: {\em paths}  $\psi (t/T)$ are in the adiabatic regime if  $F(T)\ge 0.999$.  (b). $T$ and $\Delta$ are the same as in (a). The black solid line shows the dynamics of $F$ for the case where no signal
is applied; the red dashed line ( the
blue dotted line) shows the fidelity dynamics when applying fast signals $%
2\cos^{2}(10t)$ ($2\sin^{2}(10t)$), and the dark cyan dash dot line shows $F(t)$ 
controlled by fast pulse signals with the pulse strength $s=2.0$.}
\end{figure}

Different types of fast signals work as perfect as regular rectangular
pulses~\cite{Jun15}. Fig.~$1b$ shows the fidelity dynamics by applying
different fast signals, even random signals as in Fig.~$3$. The red dashed
line shows the result by an even simpler fast signal $c(t)=2\cos^2(10t)$ and the blue
dotted line is that of $2\sin^2(10t)$. The black solid line uses
regular rectangular pulses with $s=2.0$ and $\Delta =0.08$, as a reference.

Fast signals reduce the runtime of adiabatic evolution algorithms greatly,
and keep very high fidelity $F$ particularly when the system reaches our
target--the ground state of the problem Hamiltonian $H_{P}$. Furthermore,
the runtime can be even shorten for example to half,  $T=20$, with the pules
interval $0.04$. We set the strength of pulses as $s=0,1.0,2.0$,
respectively, as in Fig.~$2$. It shows again that the adiabaticity is
greatly enhanced by increasing strengths. 

Adiabaticity can be induced from an originally fast dynamical process if pules signals are even stronger .
For example, if the signal strength $s=15$, the system wave function evolves along the adiabatic path
in the runtime $T=9$ and at the very high fidelity $F=0.999$ overlapping with the eigenstate of $H_B$, which is 17 times faster than 
the natural adiabatic process where the runtime $T_0=160$.  Numerical analysis shows that if we are allowed to increase the strength at will,
the runtime $T$ can be as fast as we wish. Other examples are, if $ s=1.0,5, 30$, $T\approx70, 23,5.0,$
respectively. 
\begin{figure}[tbp]
\includegraphics[width=0.9\columnwidth, clip]{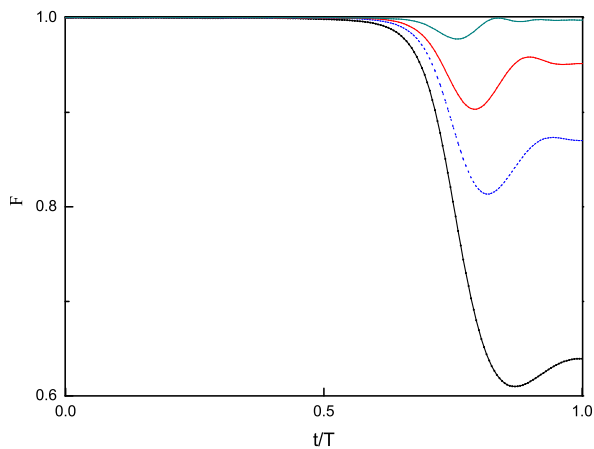}
\caption{(Color online) Dynamics of fidelity $F$ with the runtime
 $T=20$ and the intervals  $\Delta =0.04$. The black solid line shows the dynamics
for $s=0$; the blue dotted line for $s=1.0$; the red
dashed line shows the result for $s=2.0$ and the dark cyan dash dot line
for $s=5.0$.}
\end{figure}


\emph{Randomized Trotter formula and implementation of the algorithm in
trapped ions.}-- We now discuss the feasibility to experimentally implement
our algorithm on an ion trap quantum information processor. 

In general, the EC3 problem Hamiltonian is supposedly stored in a Oracle and
is called when it is needed. In order to perform experimental demonstration of our
algorithm, here we simulate the $4$-bit EC$3$ Hamiltonian with trapped
ions. We first write the problem Hamiltonian explicitly in the qubit space,
\begin{eqnarray}
H_{P} &=&\frac{3}{8}\left( \sigma _{z}^{1}\sigma _{z}^{2}\sigma
_{z}^{3}+\sigma _{z}^{1}\sigma _{z}^{2}\sigma _{z}^{4}+\sigma _{z}^{2}\sigma
_{z}^{3}\sigma _{z}^{4}\right) +  \notag \\
&&\frac{1}{4}\left( \sigma _{z}^{1}\sigma _{z}^{2}+\sigma _{z}^{2}\sigma
_{z}^{3}+\sigma _{z}^{2}\sigma _{z}^{4}\right) +  \notag \\
&&\frac{1}{8}\left( \sigma _{z}^{1}\sigma _{z}^{3}+\sigma _{z}^{1}\sigma
_{z}^{4}+\sigma _{z}^{3}\sigma _{z}^{4}\right) -  \notag \\
&&\frac{1}{4}\left( \sigma _{z}^{1}+\sigma _{z}^{3}+\sigma _{z}^{4}\right) +%
\frac{15}{8}.
\end{eqnarray}%
Note that the Hamiltonian contains up to three-body interactions, since symmetry rules out more complicated interactions which may appear in 
multi-bit EC$3$ problems.

The time-ordering evolution operators driven by the time dependent
Hamiltonians $H\left( t\right) $ and $H_{0}(t)$ cannot be analogously
simulated by trapped ions. Therefore digital simulation has to be employed.
The standard recipe of digital simulation for adiabatic processes is the use of the Trotter formula,
as done in previous literatures Ref.~\cite{Wu}.
In what follows, we will present a \emph{randomized} Trotter formula~(RTF)
to mimic $H(t)$, which effectively combine the two processes, applying fast
signals during the dynamics and simulating $H_{0}\left( t\right) $.

The time-ordering unitary evolution operator is implemented as
\begin{equation}
U\left( k\tau \right) \approx e^{-iH_{0}\left( k\tau \right) \tau
_{k}}\cdots e^{-iH_{0}\left( 2\tau \right) \tau _{2}}e^{-iH_{0}\left( \tau
\right) \tau _{1}}  \label{RTF}
\end{equation}%
up to order $O\left( \tau ^{2}\right)$ and where $j=1,\ldots ,k$. Usually,
the evolution operator of $H_0(k\tau)$ is simulated by setting all $%
\tau_j=\tau$.

The distinctive recipe of our RTF is that we set
\begin{equation}
\tau _{j}=\left( 1+\frac{c\left( j\tau \right) }{J_{0}}\right) \tau ,
\end{equation}%
such that $e^{-iH_{0}\left( j\tau \right) \tau _{j}}=e^{-iH\left( j\tau
\right) \tau }$. The equality links two different physical operations. The
left is the simulated $H_{0}$ evolving during a short but uneven time interval $%
\tau _{j}$, and the right means a fast signal $c(j\tau )$
has been implemented, at the time instance $j\tau $, upon $H_{0}$ that \emph{%
transforms} into the dressed $H$ evolving in an even time interval $\tau $. The mathematical
equivalence implies that we can experimentally simulate $e^{-iH_{0}\left( j\tau \right)
\tau _{j}}$ instead of $e^{-iH\left( j\tau \right) \tau }$, whose simulation
ingredient is not yet known ({\em unknown} for this model but it is simple to implement $c(t)$ upon $H_0$ for 
most of systems, such as an additional magnetic fast-varying field upon spins).
 In other words, the simulation (\ref{RTF}) for $H_{0}$ becomes that of $H$,
\begin{equation}
U\left( k\tau \right) \approx e^{-iH\left( k\tau \right) \tau }\cdots
e^{-iH\left( 2\tau \right) \tau }e^{-iH\left( \tau \right) \tau }
\label{RTF1}
\end{equation}%
up to order $O\left( \tau ^{2}\right) $. 

The evolution operator of $H_{0}$
is simulated by the Trotter decomposition,
\begin{equation}
e^{-iH_{0}\left( j\tau \right) \tau _{j}}\approx e^{-iH_{B}\left(
1-j/k\right) \tau _{j}}e^{-iH_{P}\left( j/k\right) \tau _{j}}.
\end{equation}%
Experimentally, exact control of these uneven time intervals $\tau _{j}$
might not be easy. Therefore, the easiest way for experimentalists is to
assign random values to these intervals $\tau _{j}$. This is equivalent to
employ random fast signals $c(j\tau )$, which has shown the same excellent control
quality as that of other fast
signals ~\cite{Jun14,Jun15}.

We set the runtime $T=20$ and let $\tau _{j}$ change randomly in the range $%
[2.0\tau ,3.0\tau ]$ and $[4.0\tau ,8.0\tau ]$ respectively, and perform
simulation. Fig.~$3$ shows the results  and compares them with regular pulses.
It is clear that random fast signals work as perfect as regular pulses. When
the variation range of $\tau _{j}$ is larger, the enhancement to
adiabaticity is even better than that of fixed $\tau _{j}$'s, and evolvs on
the same adiabatic path as that of the adiabatic reference where $T_{0}=160$.
\begin{figure}[tbp]
\includegraphics[width=0.9\columnwidth, clip]{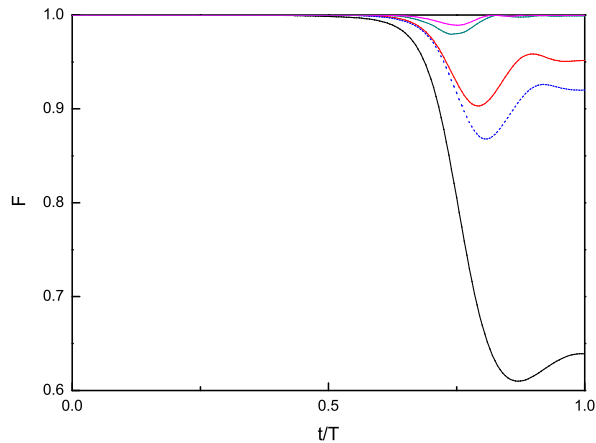}
\caption{(Color online) Dynamics of $F(t)$ with the evolution
time $T=20$. The black solid line shows $F$  without signal control
; the red dashed line is $F$ controlled by a sequence of fast pulses with $s=2.0$ and $%
\Delta =0.04$; the blue dotted
and the dark cyan dash dot lines show the dynamics of $F$ obtained by the simulated Eq.~($11$), 
where $\protect\tau _{j}$ varies randomly in the range $[2.0\protect\tau ,3.0\protect\tau ]$ and $[4.0\protect%
\tau ,8.0\protect\tau ]$, respectively. The magenta short dashed line is 
the $T_{0}=160$ reference for the adiabatic regime. }
\end{figure}

Now we come to discuss the experimental implementation on trapped ions. It
is clear that we need only to implement the slices $e^{-iH_{B}\left(
1-j/k\right) \tau _{j}}e^{-iH_{P}\left( j/k\right) \tau _{j}}$ and repeat
them to perform the evolution operator $U(k\tau )$ until $k\tau =T$. $H_{B}$
is a simple single-qubit Hamiltonian and can be implemented on most of
sophistic quantum devices, including trapped ions. It is a challenge for
quantum devices to implement three or more body interactions. Fortunately,
trapped ions do not have this difficulty. Consider tensor products of Pauli
matrices in the form of $A=\prod_{m=1}^{n}\sigma _{\alpha }^{m}$ with $%
\sigma _{\alpha }^{m}\in \left\{ 1,\sigma _{x}^{m},\sigma _{z}^{m}\right\} $%
. These products can be implemented efficiently with the M\o lmer-S\o %
rensen~(MS) scheme~\cite{Muller} on trapped ions,
\begin{equation}
e^{i\phi \sigma _{x}^{1}\otimes \sigma _{x}^{2}\otimes \cdots \otimes \sigma
_{x}^{n}}=U_{MS}\left( -\pi /2,0\right) U_{anc}\left( \phi \right)
U_{MS}\left( \pi /2,0\right) ,
\end{equation}%
where the exponential is implemented by two MS gates to the $n$ system ions
and an ancilla qubit~(no.~$0$), $U_{MS}\left( \theta ,\varphi \right) =\exp
\left( -i\frac{\theta }{4}\left( \cos \varphi S_{x}+\sin \varphi
S_{y}\right) ^{2}\right) ,$ and $S_{x,y}=\sum_{j=0}^{n}\sigma _{x,y}^{j}$. $%
U_{anc}\left( \phi \right) $ is defined as when $n$ is odd, $U_{anc}\left(
\phi \right) =e^{-i\phi \sigma _{0}^{y}}$ for\ $n=4m+1$, and $U_{anc}\left(
\phi \right) =e^{i\phi \sigma _{0}^{y}}$ for\ $n=4m-1$, and when $n$ is
even, $U_{anc}\left( \phi \right) =e^{i\phi \sigma _{0}^{z}}$ for\ $n=4m$, $%
U_{anc}\left( \phi \right) =e^{-i\phi \sigma _{0}^{z}}$ for\ $n=4m-2$.

\emph{Discussion.}-- A short runtime is of crucial importance for adiabatic
quantum algorithms to achieve polynomial time speedups over their classical
counterpart, because it is difficult to keep quantumness of a system for
long time in presence of noisy environment. In this paper, we propose an
adiabatic quantum algorithm assisted with fast signal and show that by
applying a sequence of fast signals, the runtime in the adiabatic quantum
computing can be greatly reduced. This technique has practical interest in
the physical implementation of adiabatic quantum algorithms. We applied this
approach to solving the EC$3$ problem and discuss the feasibility to implement it on
trapped ions. We introduce a randomized Trotter formula which effectively
implements effects of fast signals upon the original Hamiltonian, which,
as we show, can be implemented efficiently on a trapped ion system.

\acknowledgments

This work was supported by the National Nature Science Foundation of
China~(Grants No.~11275145), \textquotedblleft the Fundamental Research
Funds for the Central Universities\textquotedblright\ of China, the Basque
Government~(grant IT472-10) and the Spanish
MICINN~(No.~FIS2012-36673-C03-03).

\clearpage


{\center{{\large{\bf{Strength for Speed: Expedited Adiabatic Process}}}}}
\bigskip
{\center{ Lian-Ao Wu}}
\bigskip

{\center{\em{Department of Theoretical Physics and History of Science, The Basque
Country University~(EHU/UPV), P. O. Box 644, 48080 Bilbao,
and IKERBASQUE, Basque Foundation for Science, 48011 Bilbao, Spain}}}

\bigskip

%
\textbf{
Adiabatic theorem ensures that a quantum system remains on its adiabatic path: an instantaneous eigenstate of the driving
Hamiltonian, and provides the theoretical basis of adiabatic quantum information processor (AQIP). A functional large scale AQIP
is one of the most promising candidates for the ultimate universal quantum computer~\cite{Ari}, where each algorithm is designed to run through 
a specifically programmed adiabatic path. 
While adiabatic processor is claimed to have great advantages~\cite{Farhi}, recent experiments on the D-Wave Two system
found no evidence of quantum speedup~\cite{lidar1}. One of main reasons for the dysfunction could be that the runtime is so long that before the end of adiabatic quantum path, decoherence has completely ruined the quantumness. Speedup of programmed adiabatic paths is crucial in realization of practical large scale quantum computation.  Here we find that a time scaling transformation can transfer strength of 
 the driving Hamiltonian into speed of the adiabatic process. We prove rigorously that 
if it has strong enough strength, a generic non-adiabatic Hamiltonian can become an adiabatic Hamiltonian in {\em the scaling time domain}. This offers 
in principle unlimited speedup for passing through adiabatic paths.
}
 
Quantum adiabatic theorem can be formulated as follows: in the adiabatic regime, the Schr\"{o}dinger equation
\begin{equation}\label{eq1}
H_0(\frac{t}{T_0})\psi(t)=i\partial_t \psi(t)
\end{equation}
has the solution $\psi(t)=e^{i\theta_n(t)}| E_n(t)\rangle$ if initially $\psi(0)=| E_n(0)\rangle$, where $| E_n(t)\rangle$'s are instantaneous eigenstates of $H_0(\frac{t}{T_0})$. 
Here $e^{i\theta_n(t)}$ is a phase factor and $T_0$ is the total evolution time or runtime characterizing the duration needed for $H_0(\frac{t}{T_0})$ to become adiabatic, which is determined by the standard adiabatic conditions.

Consider a Hamiltonian $H(\frac{t}{T})=JH_0(\frac{t}{T})$, where $J>1$ and $T<T_0$ such that $H_0(\frac{t}{T})$ is not adiabatic. The Hamiltonian $H$ has stronger strength than $H_0$. Physically, for instance the strong strength of the Hamiltonian of a spin in a magnetic field can be made by increasing the amplitude of the magnetic field.  
The Schr\"{o}dinger equation for $H$ is $JH_0(\frac{t}{T})\psi'(t)=i\partial_t \psi'(t)$. 

We now apply a time scaling transformation $\tau=Jt$ such that the Schr\"{o}dinger equation of $H$ becomes 
\begin{equation}\label{eq2}
H_0(\frac{\tau}{JT})\psi'(\tau/J)=i\partial_\tau \psi'(\tau/J),
\end{equation}
in the scaling time $\tau$ domain, where the Hamiltonian and time derivative $\partial_\tau$ have the exact same profile as those in the Schr\"{o}dinger equation (\ref{eq1})  {\em when} $JT=T_0$. Therefore, the Schr\"{o}dinger equation of $H$ in the scaling time domain is {\em identical} to that of $H_0$ in the real time $t$ domain, such that their solutions are identical,   
\begin{equation}\label{eq3}
\psi'(\tau/J)\equiv \psi(\tau)=e^{i\theta_n(\tau)}| E_n(\tau)\rangle.
\end{equation}
This states a {\em quantum adiabatic theorem in the scaling time domain}, and is also a formal proof that the runtime of an adiabatic quantum process can be {\em reduced $J$ times} or decreases from $T_0$ to $T=T_0/J$.  Note that the standard adiabatic conditions remain unchanged in the scaling time domain, except $t$ being replaced by $\tau$.

In analogy with the normal adiabatic theorem for adiabatic processes,  the quantum adiabatic theorem in the scaling time domain provides the basis of expedited adiabatic processes.

An example is the adiabatic algorithm in Ref.~\cite{Farhi}, where 
\begin{equation}\label{eq4}
H(t/T)=J[(1-t/T)H_B+(t/T)H_P].
\end{equation}
The expedited adiabatic process runs on the adiabatic path in the scaling time domain: $| E_0(t/J)\rangle$ (note that $\tau$ is replaced by $t/J$), where the initial state is ground state of $H_B$ and the final state is ground state of $H_P$ encoding solution to the problem of interest. The evolution from $H_B$ to $H_P$ needs only $T=T_0/J$, for example, as a limit the runtime of an adiabatic algorithm could be $T\approx 0$ if $J\rightarrow\infty$.


The quantum adiabatic theorem in the scaling time domain clearly suggests the strategy of experimentally implementing an expedited adiabatic processes: {\em simply pushing the strength $J$ to its upper bound} (and combining with fast signal proposal to further improve the adiabaticity as in Ref. \cite{Jun14, Wang}). The quantum adiabatic theorem offers 
in principle unlimited speedup (non relativistic) for passing through adiabatic paths when the strength $J$ is very strong.


\bigskip

 \textbf{Acknowledgments}

\bigskip
This work was supported by the Basque
Government grant (IT472-10) and the Spanish
MICINN (No.~FIS2012-36673-C03-03).




\clearpage


\begin{thebibliography}{99}
\bibitem{Born} M. Born and V. Fock, Z. Phys. 51, 165~(1928).

\bibitem{Messiah} A. Messiah, Quantum mechanics~(Amsterdam: North-Holland,
1962).

\bibitem{Farhi} E. Farhi, J. Goldstone, S. Gutmann, J. Lapan, A. Lundgren,
and D. Preda, Science 292, 472~(2001).

\bibitem{Childs} A. M. Childs, E. Farhi, and J. Preskill, \pra{65},
012322~(2001).

\bibitem{Aharonov} D. Aharonov, et.al, arXiv:quant-ph/0405098.

\bibitem{Zanardi} P. Zanardi and M. Rasetti, Phys. Lett. A, {264}, 94~(1999).

\bibitem{Carollo} A. C. M. Carollo, V. Vedral, arXiv:quant-ph/0504205.

\bibitem{Berry} M. Berry, Proc. R. Soc. Lond. A 392, 45~(1984).

\bibitem{Zee} F. Wilczek and A. Zee, Phys. Rev. Lett. {52}, 2111~(1984).

\bibitem{Barry} K. -P. Marzlin and B. C. Sanders, Phys. Rev. Lett. {93},
160408~(2004).

\bibitem{shor} P.~W. Shor, in Proceedings of the Symposium on the
Foundations of Computer Science, 1994, Los Alamitos, California~(IEEE
Computer Society Press, New York, 1994), pp.~124--134.

\bibitem{grover} L.~K. Grover, Phys. Rev. Lett. \textbf{79}, 325~(1997).

\bibitem{Fey} R. Feynman, Inter. J. Theor. Phys. \textbf{21}, 467~(1982).

\bibitem{lidar1} T. F. R\o nnow, et. al, Science \textbf{345}, 420~(2014).

\bibitem{farhi1} E. Farhi, J. Goldstone, S. Gutmann, e-print
quant-ph/0007071v1~(2000).

\bibitem{Jun14} Jun Jing, L.-A Wu, T. Yu, J.~Q. You, Z.-M. Wang, and L.
Garcia, Phys. Rev. A \textbf{89}, 032110~(2014). See also the attached separated file by Lian-Ao Wu.

\bibitem{Wu} L. -A. Wu, M. S. Byrd and D. A. Lidar, Phys. Rev. Lett. 8{9},
057904~(2002).

\bibitem{Jun15} Jun Jing, L.-A Wu, M. Byrd, J. Q. You, T. Yu and Z.-M. Wang,

\bibitem{Muller} M. M\"{u}ller, K. Hammerer, Y. L. Zhou, C. F. Roos and P.
Zoller, New. J. Phys. {13}, 085007~(2011).

\bibitem{MS} K. M\o lmer and A. S\o rensen, Phys. Rev. Lett. 82, 1835~(1999).
\end{thebibliography}

\begin{thebibliography}{99}

\bibitem{Ari} A. Mizel, D. A. Lidar, and M. W. Mitchell, Simple proof of equivalence between adiabatic quantum computation and the circuit model, Phys. Rev. Lett. {\bf 99}, 070502 (2007).

\bibitem{Farhi} E. Farhi, J. Goldstone, S. Gutmann, J. Lapan, A. Lundgren,
and D. Preda, A quantum adiabatic evolution algorithm applied to random instances of an NP-complete problem, Science 292, 472~(2001).

\bibitem{lidar1} T. F. R\o nnow, Z. Wang, J. Job, S.V. Isakov, D. Wecker, J.M. Martinis, D.A. Lidar, and M. Troyer,  Defining and Detecting Quantum Speedup, Science \textbf{345}, 420~(2014).

\bibitem{Jun14} Jun Jing, L.-A Wu, T. Yu, J.~Q. You, Z.-M. Wang, and L.
Garcia, One-component dynamical equation and noise-induced adiabaticity, Phys. Rev. A \textbf{89}, 032110~(2014).

\bibitem{Wang} H. Wang and L. -A. Wu,  Fast quantum algorithm for EC3 problem with trapped ions, arXiv:1412.1722 

\end{thebibliography}
\end{document}